# General Impedance Modeling for Modular Multilevel Converter with Grid-forming and Grid-following Control

Chu Sun, *Member, IEEE*, Fei Zhang, *Member, IEEE,* Huafeng Xiao, *Senior Member, IEEE*, Na Wang, *Member, IEEE,* and Jikai Chen, *Member, IEEE*

*Abstract*--Modular multilevel converter (MMC) has complex topology, control architecture and broadband harmonic spectrum. For this, linear-time-periodic (LTP) theory, covering multi-harmonic coupling relations, has been adopted for MMC impedance modeling recently. However, the existing MMC impedance models usually lack explicit expressions and general modeling procedure for different control strategies. To this end, this paper proposes a general impedance modeling procedure applicable to various power converters with grid-forming and grid-following control strategies. The modeling is based on a unified representation of MMC circuit as the input and output relation between the voltage or current on the AC side and the exerted modulation index, while the control part vice versa, thereby interconnected as closed-loop feedback. With each part expressed as transfer functions, the final impedance model keeps the explicit form of harmonic transfer function matrix, making it convenient to directly observe and analyze the influence of each part individually. Thereby the submodule capacitance is found as the main cause of difference between MMC impedance compared to two-level converter, which will get closer as the capacitance increases. Effectiveness and generality of the impedance modeling method is demonstrated through comprehensive comparison with impedance scanning using electromagnetic transient simulation.

*Index Terms*-- Modular multilevel converter, impedance model, grid-forming, grid-following, linear time periodic theory.

## I. Introduction

IN the past decade, harmonic stability issues caused by the large-scale integration of power electronic converters in power system, have become increasingly prominent [1]. To address this issue, impedance modeling methods for stability analysis have been widely adopted [2]. To cover broadband harmonics for converter such as MMC, and accommodate single-phase or unbalanced operating conditions, a category of modeling methods based on linear-time-periodic (LTP) theory has been proposed [3]. This theory decomposes variables as multiple periodical signals in frequency domain and established the input-output relation as impedance matrix, with the harmonic orders flexibly selectable [4]. LTP theory also spawned multiple modeling methods such as dynamic phasors, multi-harmonic linearization, harmonic state-space (HSS), and harmonic transfer functions (HTF), etc.

From the historical perspective, M. Madrigal and A. Wood, et al. proposed extended harmonic modeling and HSS models for AC-DC converters such as STATCOM and thyristor-rectified HVDC, without considering the control loops [5-6]. J. Sun et al. proposed harmonic linearization method in the stationary reference frame to model AC-DC converters based on positive and negative sequence impedances, but their initial work neglected the couplings between these sequences and is difficult to cover multi-harmonic coupling [7]. Their recent work was extended to higher-order harmonic linearization and considered sequence-couplings, with derivation performed based on Toeplitz matrix operation, which is similar to HSS method in terms of theoretical foundation [8]. L. Harnefors et al. proposed impedance model in complex coordinate system at fundamental frequency [9], and X. Wang et al. further developed HSS method in stationary coordinate system, including both converter circuit and control [10]. C. Zhang and A. Rygg et al. revealed the equivalence and transformation of impedance in different coordinate systems at fundamental frequency and proposed harmonic transfer function (HTF) method, but mainly used for two-level converters [11-12]. The relations between DQ-domain impedance, sequence impedance, and HSS impedance models is revealed in [13], which points out HSS impedance is a high-dimensional generalization of sequence impedance model. Y. Li proposed frequency-domain transformation relations between various coordinate systems, laying the foundation for transformation of impedance models with high harmonic orders, but did not derive impedance models for complex converters such as MMC [14].

LTP theory-based modeling approach has been applied to various converter topologies and control strategies. MMC impedance model based on dynamic phasor approach is developed in [15] and [16], which is essentially in DQ complex coordinate system, with the final model formed as a high-dimensional matrix. Multi-input multi-output impedance model

This work has been supported in part by National Natural Science Foundation of China (52077030).
Chu Sun is with the School of Electrical Engineering and Computer Science, KTH Royal Institute of Technology, Stockholm, 11428, Sweden (e-mail: chus@kth.se).
Fei Zhang and Huafeng Xiao are with the School of Electrical Engineering, Southeast University, Nanjing, 210096, China (e-mail: zhangfei@seu.edu.cn).
Na Wang is with the School of Automation Science and Electrical Engineering, Beihang University, Beijing, 100191, China.
Jikai Chen is with the Key Laboratory of Modern Power System Simulation and Control & Renewable Energy Technology, Ministry of Education, Northeast Electric Power University, Jilin 132012, China (e-mail: chenjikai@neepu.edu.cn).

for MMC in wind farm is developed in [17] based on HSS method, which can be applied to asymmetric conditions. These models are mainly for relatively simple grid-following control. For more complex control strategies, such as grid-following DC voltage or grid-forming droop control, impedance models have rarely been reported. DQ models of MMC with open-loop and closed-loop V/f control at fundamental-frequency coordinate system are derived in [18]. MMC impedance models with different types of grid-forming control are developed in [19-21] based on HSS method. T. Huang proposed sequence-impedance model for MMC with DC voltage control [22], focusing on analyzing the relationship between the DC-side voltage and AC-side variables. Though multiple models are derived for MMC, one major issue is the modeling process is not standardized for different control types, which limits their widespread acceptance.

On the other hand, a desirable impedance model should have explicit expression so that the influence of each part and the physical interpretation can be analyzed directly. The impedance model derived by W. Bo et al. has explicit form, but the admittance model is restricted to two-level converters and DQ coordinate system [23]. By contrast, most MMC models derived from LTP theory are presented in implicit state-space form, with the concerned elements extracted from high-dimensional matrix, making it difficult to visually examine the impact of each part. T. Yin et al made efforts to give explicit forms of MMC impedance, but restricted to grid-following control only [24].

As reviewed above, the impedance modeling for complex converter topologies like MMC, is still unmature and not standardized. To this end, this paper proposes a general modeling for MMCs with different control types based on LTP theory. The main contributions are as follows:

(1) A general MMC impedance model and derivation process is proposed, which is applicable to different control types, and can also be extended to other converter topologies.

(2) The proposed impedance model has an explicit expression, which facilitates the observation and analysis of the influence of each part on the impedance directly.

(3) The main difference of MMC converter admittance formulation, compared with two-level converter, is the term of submodule capacitance, which will get closer with larger value.

The remainder of the article is structed as follows: Section II introduces the basics for impedance modeling, including the general modeling procedure and the topology of MMC. Section III, IV and V present the impedance models for various control types of MMC, including the commonly used grid-forming and grid-following control strategies, and extension to two-level converter. Section VI verifies the accuracy of the impedance models, and the last section gives conclusions.

## II. FUNDAMENTALS FOR MMC IMPEDANCE MODELING

### A. General Procedure for Impedance Modeling

Without considering specific AC or DC converter topology and control strategy, power electronic converter and its control system can be uniformly represented as the diagram in Fig. 1, which is divided into control plant and control algorithm, where $\Delta u$ is the input perturbation and $\Delta y$ the output variable, which can be voltage or current, respectively; $\Delta m$ is the modulation index. These variables are scalar for DC circuit, while vector for AC circuit.

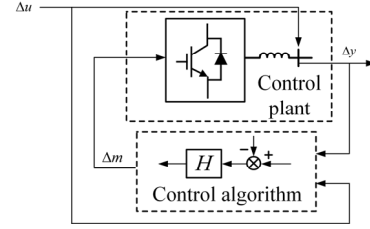

Fig. 1. General power electronic converter and control system.

The control plant consists of power converter and its filter circuit, which can be expressed as (1) by modeling each component, and interconnecting based on circuit theory.

$$Z\Delta y = C\Delta m + D\Delta u \quad (1)$$

Based on control block algebra, the control algorithm can be universally expressed as (2).

$$\Delta m = B\Delta u + F\Delta y \quad (2)$$

Combining (1) and (2), the relation between input perturbation and output response, namely, the expression of impedance or admittance, can be obtained as (3).

$$\Delta y = (Z - CF)^{-1}(CB + D)\Delta u \quad (3)$$

where $Z$, $B$, $F$, $C$ and $D$ are transfer function matrices. Therefore, the main work of impedance modeling is reduced to deriving these transfer function matrices. Among them, $Z$, $C$ and $D$ can be obtained in a standardized way based on circuit theory (nodal voltage or loop current analysis), while $B$ and $F$ can be derived based on basic rules to calculate the input-output relation of MIMO linear system, such as superposition property.

The above modeling is feasible for different AC or DC converter topologies, harmonic orders and coordinate systems, such as synchronous (DQ) or static reference frame (sequence domain PN), and extended harmonic-domain (HSS, HTF, etc.). HTF method is adopted here, which enables derivation of high-order impedance model as with conventional transfer functions. Details of LTP theory and HTF method can be found in [25-26, 12], and omitted here for brevity. In this article, MMC models are presented as example, and compared with two-level voltage-source converter (2L-VSC).

### B. Introduction to MMC Circuit

Without loss of generality, three-phase MMC with half-bridge modules is considered. A typical topology is shown in Fig. 2, where: $i_{u,x}$, $i_{L,x}$ (x=a, b, c) are the arm currents of the upper and lower bridge; $i_{g,x}$ and $u_{g,x}$ are the output current and voltage; $u_{Cu,x}$, $u_{CL,x}$ are the equivalent capacitor voltages of the upper and lower bridge arms; $L_{arm}$, $R_{arm}$ are the inductance and resistance of the bridge arm; $C_m$ is the submodule capacitance; $L_g$ is the grid-side inductance. The average model of MMC submodules can be equivalent to controlled voltage and current sources, containing equivalent capacitance $C_{eq} = C_m/N$, where $N$ is the number of modules in one bridge arm, and the switching functions of the bridge arm can be averaged by the upper and lower bridge arm modulation signals $m_{u,x}$, $m_{L,x}$.

MMC can be grid-forming or grid-following controlled,

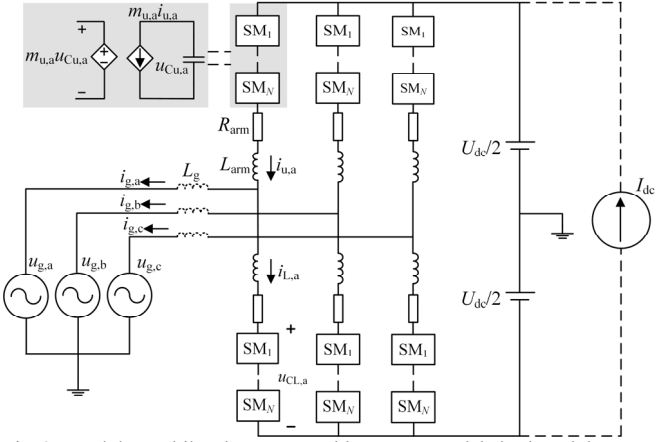

Fig. 2. Modular multilevel converter with average-modeled submodule.

depending on which the DC side can be equivalent to a constant DC voltage source $U_{dc}$ or a constant DC current source $I_{dc}$. The grid-side impedance is ignored first, and can be considered in series with the derived MMC impedance if included.

## III. Admittance Modeling of Grid-forming MMC

The diagram of MMC grid-forming control is illustrated in Fig. 3, which can also cover constant V/f control with the outer control blocks in Fig. 3(a) as zero [27]. The modeling process starts from the converter circuit, then to the control part, and is finally assembled as a whole. For simplicity but without loss of generality, low-pass filters in voltage and current measurement, and the delay in modulation, are ignored first.

### A. Modeling of MMC Main Circuit

From the equivalent average model of arm modules in Fig. 2 and the modulation index transformation relationship, LTP model of the arm capacitor voltage is obtained as:

$$\Delta \boldsymbol{u}_{C}^{ltp} = \boldsymbol{Z}_{Ceq}(\boldsymbol{I}_{s}^{ltp}\Delta \boldsymbol{m}^{ltp} + \boldsymbol{M}_{i}^{ltp}\Delta \boldsymbol{i}^{ltp}) \quad (4)$$

where, as denoted in (5)-(8), $\boldsymbol{u}_{C}^{ltp}$, $\Delta \boldsymbol{i}^{ltp}$, and $\Delta \boldsymbol{m}^{ltp}$ are the LTP representation, in the form of Toeplitz matrices, of the decomposed exponential periodic signals of capacitor voltages, common-mode and differential-mode currents, and upper and lower arm modulation indices, respectively. $h$ is the highest order considered in the harmonic spectrum. $\boldsymbol{Z}_{Ceq}$, $\boldsymbol{I}_{s}^{ltp}$, and $\boldsymbol{M}_{i}^{ltp}$ are LTP representation of the arm capacitance, arm current, and current-related modulation index, respectively. The subscript 'u' and 'l' (or 'L') indicate respectively the variables associated with the upper and lower arm.

$$\boldsymbol{u}_{C}^{ltp}(s) = \left[\boldsymbol{u}_{Cu}^{ltp}, \boldsymbol{u}_{Cl}^{ltp}\right]^{T}, \boldsymbol{u}_{Cu}^{ltp}(s) = \left[\boldsymbol{u}_{Cua}^{ltp}, \boldsymbol{u}_{Cub}^{ltp}, \boldsymbol{u}_{Cuc}^{ltp}\right]^{T}$$

$$\boldsymbol{u}_{Cl}^{ltp}(s) = \left[\boldsymbol{u}_{Cla}^{ltp}, \boldsymbol{u}_{Clb}^{ltp}, \boldsymbol{u}_{Clc}^{ltp}\right]^{T}$$

$$\boldsymbol{u}_{Cla}^{ltp}(s) = \left[u_{a,-h}, \cdots, u_{a,0}, \cdots, u_{a,+h}\right]^{T}$$

$$\boldsymbol{i}^{ltp}(s) = \left[\boldsymbol{i}_{c}^{ltp}, \boldsymbol{i}_{g}^{ltp}\right]^{T}, \boldsymbol{i}_{c}^{ltp}(s) = \left[\boldsymbol{i}_{ca}^{ltp}, \boldsymbol{i}_{cb}^{ltp}, \boldsymbol{i}_{cc}^{ltp}\right]^{T} \quad (5)$$

$$\boldsymbol{i}_{g}^{ltp}(s) = \left[\boldsymbol{i}_{ga}^{ltp}, \boldsymbol{i}_{gb}^{ltp}, \boldsymbol{i}_{gc}^{ltp}\right]^{T}$$

$$\boldsymbol{m}^{ltp} = \begin{bmatrix} \boldsymbol{m}_{u}^{ltp} \\ \boldsymbol{m}_{L}^{ltp} \end{bmatrix} = \begin{bmatrix} -\boldsymbol{m}_{1}^{ltp} - \boldsymbol{m}_{2}^{ltp} \\ \boldsymbol{m}_{1}^{ltp} - \boldsymbol{m}_{2}^{ltp} \end{bmatrix}/2$$

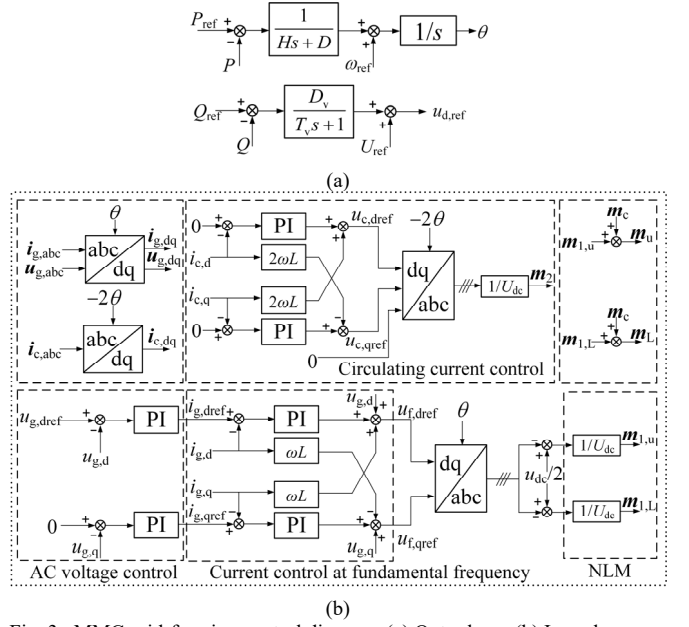

Fig. 3. MMC grid-forming control diagram. (a) Outer loop. (b) Inner loop.

$$\boldsymbol{Z}_{Ceq} = \left[\boldsymbol{Z}_{C}, \boldsymbol{Z}_{C}\right]^{T}, \boldsymbol{Z}_{C} = \text{blkdiag}\{\boldsymbol{Z}_{Cx}\}, x=a,b,c \quad (6)$$

$$\boldsymbol{Z}_{Cx} = \text{diag}\{C_{arm}(s-jh\omega_{0}), \cdots, C_{arm}s, \cdots, C_{arm}(s+jh\omega_{0})\}^{-1} \quad (7)$$

$$\boldsymbol{I}_{s}^{ltp} = \begin{bmatrix} \boldsymbol{I}_{c}^{ltp} + \boldsymbol{I}_{g}^{ltp}/2 \\ & \boldsymbol{I}_{c}^{ltp} - \boldsymbol{I}_{g}^{ltp}/2 \end{bmatrix}, \boldsymbol{M}_{i}^{ltp} = \begin{bmatrix} \boldsymbol{M}_{u}^{ltp} & \boldsymbol{M}_{u}^{ltp}/2 \\ \boldsymbol{M}_{L}^{ltp} & -\boldsymbol{M}_{L}^{ltp}/2 \end{bmatrix} \quad (8)$$

Similarly, LTP model of MMC arm inductor is:

$$\boldsymbol{Z}_{Larm}\Delta \boldsymbol{i}^{ltp} = -(\boldsymbol{E}_{t}\Delta \boldsymbol{u}_{g}^{ltp} + \boldsymbol{U}_{C}^{ltp}\Delta \boldsymbol{m}^{ltp} + \boldsymbol{M}_{v}^{ltp}\Delta \boldsymbol{u}_{C}^{ltp}) \quad (9)$$

where $\boldsymbol{u}_{g}^{ltp}$ is grid-side voltage, $\boldsymbol{u}_{C}^{ltp}$ is the phase voltage of the upper and lower arms of MMC, $\boldsymbol{Z}_{Larm}$ is the arm impedance, and $\boldsymbol{M}_{v}$ is the modulation index matrix for voltage conversion:

$$\boldsymbol{Z}_{Larm} = \left[\boldsymbol{Z}_{L}, \boldsymbol{Z}_{L}\right]^{T}, \boldsymbol{Z}_{L} = \text{blkdiag}\{\boldsymbol{Z}_{Lx}\}, x=a,b,c \quad (10)$$

$$\boldsymbol{Z}_{Lx} = \text{diag}\{L_{arm}(s-jk\omega_{0}), \cdots, L_{arm}s, \cdots, L_{arm}(s+jk\omega_{0})\} + R_{arm}\boldsymbol{I}_{2h+1} \quad (11)$$

$$\boldsymbol{U}_{C}^{ltp} = \begin{bmatrix} -\dfrac{\boldsymbol{U}_{Cux}^{ltp}}{2} & -\dfrac{\boldsymbol{U}_{Clx}^{ltp}}{2} \\ -\boldsymbol{U}_{Cux}^{ltp} & \boldsymbol{U}_{Clx}^{ltp} \end{bmatrix}, \boldsymbol{M}_{v}^{ltp} = \begin{bmatrix} -\dfrac{\boldsymbol{M}_{ux}^{ltp}}{2} & -\dfrac{\boldsymbol{M}_{lx}^{ltp}}{2} \\ -\boldsymbol{M}_{ux}^{ltp} & \boldsymbol{M}_{lx}^{ltp} \end{bmatrix}, \boldsymbol{E}_{t} = \begin{bmatrix} 0 \\ 2\boldsymbol{I}_{3(2h+1)} \end{bmatrix} \quad (12)$$

Substituting (9) into (4), we obtain the model of main circuit, which is consistent with the general model (1).

$$(\boldsymbol{Z}_{Larm} + \boldsymbol{M}_{v}^{ltp}\boldsymbol{Z}_{Ceq}\boldsymbol{M}_{i}^{ltp})\Delta \boldsymbol{i}^{ltp} + \boldsymbol{E}_{t}\Delta \boldsymbol{u}_{g}^{ltp} = -(\boldsymbol{U}_{C}^{ltp} + \boldsymbol{M}_{v}^{ltp}\boldsymbol{Z}_{Ceq}\boldsymbol{I}_{s}^{ltp})\Delta \boldsymbol{m}^{ltp} \quad (13)$$

### B. Outer Control Loop and Synchronization

The small-signal model of the active power of MMC is:

$$\Delta P^{ltp} = 1.5(\boldsymbol{U}_{g,d}^{ltp}\Delta \boldsymbol{i}_{g,d}^{ltp} + \boldsymbol{I}_{g,d}^{ltp}\Delta \boldsymbol{u}_{g,d}^{ltp} + \boldsymbol{U}_{g,q}^{ltp}\Delta \boldsymbol{i}_{g,q}^{ltp} + \boldsymbol{I}_{g,q}^{ltp}\Delta \boldsymbol{u}_{g,q}^{ltp}) \quad (14)$$

For grid-forming control diagram in Fig. 3(a), the small-signal model of the internal phase is given by (15), where the transfer function from active power to phase is (16), and the range of harmonic order $k$ considered is $[-h, h]$. $H$ and $D$ are the values of virtual inertial and damping, respectively.

$$\Delta \boldsymbol{\theta}^{ltp} = -\boldsymbol{G}_{APC}^{ltp} \Delta \boldsymbol{P}^{ltp} \tag{15}$$

$$\boldsymbol{G}_{APC}^{ltp} = \text{diag}\left(\frac{1}{(s+jk\omega_0)[H(s+jk\omega_0)+D]}\right) \tag{16}$$

From this, phase angle can be derived as (17), where the intermediate valuables are given by (18) and (19).

$$\Delta \boldsymbol{\theta}^{ltp} = \boldsymbol{G}_{ti}\Delta \boldsymbol{i}_{g,dq}^{ltp} + \boldsymbol{G}_{tu}\Delta \boldsymbol{u}_{g,dq}^{ltp} \tag{17}$$

$$\boldsymbol{G}_{ti} = -1.5\boldsymbol{G}_{APC}^{ltp}\left[\boldsymbol{U}_d^{ltp}, \boldsymbol{U}_q^{ltp}\right], \boldsymbol{G}_{tu} = -1.5\boldsymbol{G}_{APC}^{ltp}\left[\boldsymbol{I}_d^{ltp}, \boldsymbol{I}_q^{ltp}\right] \tag{18}$$

$$\Delta \boldsymbol{i}_{g,dq}^{ltp} = \left[\Delta \boldsymbol{i}_{g,d}^{ltp}, \Delta \boldsymbol{i}_{g,q}^{ltp}\right]^T, \Delta \boldsymbol{u}_{g,dq}^{ltp} = \left[\Delta \boldsymbol{u}_{g,d}^{ltp}, \Delta \boldsymbol{u}_{g,q}^{ltp}\right]^T \tag{19}$$

The small-signal models of output reactive power and voltage amplitude reference are denoted by (20) and (21), respectively, where $D_v$ is the Q-V droop gain, and $T_v$ is the associated time constant in low-pass filter.

$$\Delta \boldsymbol{Q}^{ltp} = -1.5(\boldsymbol{U}_{g,q}^{ltp}\Delta \boldsymbol{i}_{g,d}^{ltp} + \boldsymbol{I}_{g,d}^{ltp}\Delta \boldsymbol{u}_{g,q}^{ltp} - \boldsymbol{U}_{g,d}^{ltp}\Delta \boldsymbol{i}_{g,q}^{ltp} - \boldsymbol{I}_{g,q}^{ltp}\Delta \boldsymbol{u}_{g,d}^{ltp}) \tag{20}$$

$$\Delta \boldsymbol{U}_{ref}^{ltp} = -\boldsymbol{G}_{AVC}^{ltp}\Delta \boldsymbol{Q}^{ltp}, \boldsymbol{G}_{AVC}^{ltp} = \text{diag}\left(\frac{D_v}{T_v(s+jk\omega_0)+1}\right) \tag{21}$$

The small-signal model of voltage amplitude reference value is shown below.

$$\Delta \boldsymbol{U}_{ref}^{ltp} = \boldsymbol{G}_{ui,1}\Delta \boldsymbol{i}_{g,dq}^{ltp} + \boldsymbol{G}_{uu,1}\Delta \boldsymbol{u}_{g,dq}^{ltp} \tag{22}$$

$$\boldsymbol{G}_{ui,1} = -1.5\boldsymbol{G}_{APC}^{ltp}\left[\boldsymbol{U}_{g,q}^{ltp}, -\boldsymbol{U}_{g,d}^{ltp}\right], \boldsymbol{G}_{uu,1} = -1.5\boldsymbol{G}_{APC}^{ltp}\left[-\boldsymbol{I}_{g,q}^{ltp}, \boldsymbol{I}_{g,d}^{ltp}\right] \tag{23}$$

Expanding to dq variables, the voltage reference can then be rewritten as:

$$\Delta \boldsymbol{u}_{ref,dq}^{ltp} = \left[\Delta \boldsymbol{U}_{ref}^{ltp}, \boldsymbol{0}\right]^T = \boldsymbol{G}_{ui}\Delta \boldsymbol{i}_{g,dq}^{ltp} + \boldsymbol{G}_{uu}\Delta \boldsymbol{u}_{g,dq}^{ltp} \tag{24}$$

$$\boldsymbol{G}_{ui} = \left[\boldsymbol{G}_{ui,1}, \boldsymbol{0}\right]^T, \boldsymbol{G}_{uu} = \left[\boldsymbol{G}_{uu,1}, \boldsymbol{0}\right]^T \tag{25}$$

### C. Coordinate Transformation

Taking grid-side coordinate system as the reference system, LTP model of Park transform for fundamental and double frequency coordinate system is:

$$\boldsymbol{T}_{abc/dq}^{ltp(\theta)} = \frac{2}{3}\begin{bmatrix} \boldsymbol{T}_{\cos a}^{ltp} & \boldsymbol{T}_{\cos b}^{ltp} & \boldsymbol{T}_{\cos c}^{ltp} \\ -\boldsymbol{T}_{\sin a}^{ltp} & -\boldsymbol{T}_{\sin b}^{ltp} & -\boldsymbol{T}_{\sin c}^{ltp} \end{bmatrix} = \begin{bmatrix} \boldsymbol{T}_{\cos}^{ltp} \\ \boldsymbol{T}_{\sin}^{ltp} \end{bmatrix}$$

$$\boldsymbol{T}_{abc/dq}^{ltp(2\theta)} = \frac{2}{3}\begin{bmatrix} \boldsymbol{T}_{\cos 2a}^{ltp} & \boldsymbol{T}_{\cos 2b}^{ltp} & \boldsymbol{T}_{\cos 2c}^{ltp} \\ -\boldsymbol{T}_{\sin 2a}^{ltp} & -\boldsymbol{T}_{\sin 2b}^{ltp} & -\boldsymbol{T}_{\sin 2c}^{ltp} \end{bmatrix} \tag{26}$$

where each submatrix are Toeplitz matrices, for example, $\boldsymbol{T}_{\cos b}^{ltp}$ and $\boldsymbol{T}_{\cos 2b}^{ltp}$ are Toeplitz matrices of $\cos(\omega_0 t - \pi/3)$ and $\cos(2\omega_0 t + 2\pi/3)$ respectively. The non-zero terms of $\boldsymbol{T}_{\cos b}^{ltp}$ are $T_{+1} = a^*/2$, $T_{-1} = a/2$ ($a = e^{j2\pi/3}$). The LTP model of inverse Park transform is the transpose of Park transform model:

$$\boldsymbol{T}_{dq/abc}^{ltp(\theta)} = \begin{bmatrix} \boldsymbol{T}_{\cos a}^{ltp} & \boldsymbol{T}_{\cos b}^{ltp} & \boldsymbol{T}_{\cos c}^{ltp} \\ -\boldsymbol{T}_{\sin a}^{ltp} & -\boldsymbol{T}_{\sin b}^{ltp} & -\boldsymbol{T}_{\sin c}^{ltp} \end{bmatrix}^T \tag{27}$$

### D. Inner Control Loop and Modulation Index Generation

The inner voltage and current control loops are shown in Fig. 3(b). With coordinate transformation considered, the AC voltage control loop is modeled as (28), where the cross-coupling term is ignored, and the transfer function matrix of voltage PI control is modeled by (29). $\Delta \boldsymbol{i}_{g,dref}^{ltp}$ and $\Delta \boldsymbol{i}_{g,qref}^{ltp}$ are Toeplitz matrices of d- and q-components of the current reference values.

$$\left[\Delta \boldsymbol{i}_{g,dref}^{ltp}, \Delta \boldsymbol{i}_{g,qref}^{ltp}\right]^T = \boldsymbol{H}_v^{ltp}(\Delta \boldsymbol{u}_{ref,dq}^{ltp} - \Delta \boldsymbol{u}_{g,dq}^{ltp} - \boldsymbol{U}_{g,qd}^{ltp}\Delta \boldsymbol{\theta}^{ltp})$$

$$\Delta \boldsymbol{u}_{g,dq}^{ltp} = \left[\Delta \boldsymbol{u}_{g,d}^{ltp}, \Delta \boldsymbol{u}_{g,q}^{ltp}\right]^T, \boldsymbol{U}_{t,qd}^{ltp} = \left[\boldsymbol{U}_{g,q}^{ltp}, -\boldsymbol{U}_{g,d}^{ltp}\right]^T \tag{28}$$

$$\boldsymbol{H}_v^{ltp} = \text{blkdiag}(\boldsymbol{G}_v^{ltp}, \boldsymbol{G}_v^{ltp}), \boldsymbol{G}_v^{ltp} = \text{diag}\left(k_{pv} + \frac{k_{iv}}{s+jk\omega_0}\right) \tag{29}$$

The fundamental current control loop is modeled as (30), while voltage feedforward control are temporarily ignored. The transfer function of the current control block is given by (31), where $K_i^{ltp}$ denotes the current cross-decoupling term.

$$\left[\Delta \boldsymbol{u}_{f,dref}^{ltp}, \Delta \boldsymbol{u}_{f,qref}^{ltp}\right]^T = -\boldsymbol{H}_i^{ltp}\boldsymbol{T}_{abc/dq}^{ltp(\theta)}\Delta \boldsymbol{i}_g^{ltp} - \boldsymbol{T}_{pll,g0}^{ltp}\boldsymbol{T}_{pll}^{ltp}\boldsymbol{T}_{\sin}^{ltp}\cdot\Delta \boldsymbol{u}_g^{ltp}$$

$$\Delta \boldsymbol{m}_{1,dq}^{ltp} = \left[\Delta \boldsymbol{u}_{f,dref}^{ltp}, \Delta \boldsymbol{u}_{f,qref}^{ltp}\right]^T / U_{dc}, \boldsymbol{U}_{f,qd}^{ltp} = \left[\boldsymbol{U}_{fq0}^{ltp}, -\boldsymbol{U}_{fd0}^{ltp}\right]^T$$

$$\boldsymbol{T}_{pll,g0}^{ltp} = \boldsymbol{H}_i^{ltp}\boldsymbol{I}_{g,qd}^{ltp} + \boldsymbol{U}_{f,qd}^{ltp}, \boldsymbol{M}_{1,qd}^{ltp} = \boldsymbol{U}_{f,qd}^{ltp}/U_{dc} \tag{30}$$

$$\Delta \boldsymbol{i}_{g,dq}^{ltp} = \left[\Delta \boldsymbol{i}_{g,d}^{ltp}, \Delta \boldsymbol{i}_{g,q}^{ltp}\right]^T, \boldsymbol{I}_{g,qd}^{ltp} = \left[\boldsymbol{I}_{g,q}^{ltp}, -\boldsymbol{I}_{g,d}^{ltp}\right]^T$$

$$\boldsymbol{H}_i^{ltp} = \begin{bmatrix} \boldsymbol{G}_i^{ltp} & -\boldsymbol{K}_i^{ltp} \\ \boldsymbol{K}_i^{ltp} & \boldsymbol{G}_i^{ltp} \end{bmatrix}, \boldsymbol{G}_i^{ltp} = \text{diag}\left(k_{pi} + \frac{k_{ii}}{s+jk\omega_0}\right) \tag{31}$$

Substituting voltage control (28) and phase angle (17) into the fundamental current control loop yields:

$$\Delta \boldsymbol{m}_{1,dq}^{ltp} = \boldsymbol{A}\Delta \boldsymbol{i}_{g,dq}^{ltp} + \boldsymbol{B}\Delta \boldsymbol{u}_{g,dq}^{ltp}$$

$$\boldsymbol{A} = \boldsymbol{G}_i^{ltp}(\boldsymbol{H}_v^{ltp}\boldsymbol{G}_{ui} - \boldsymbol{I}) - (\boldsymbol{M}_{1,qd}^{ltp} + \boldsymbol{H}_i^{ltp}\boldsymbol{I}_{g,qd}^{ltp} + \boldsymbol{H}_i^{ltp}\boldsymbol{H}_v^{ltp}\boldsymbol{U}_{t,qd}^{ltp})\boldsymbol{G}_{ti} \tag{32}$$

$$\boldsymbol{B} = \boldsymbol{G}_i^{ltp}\boldsymbol{H}_v^{ltp}(\boldsymbol{G}_{uu} - \boldsymbol{I}) - (\boldsymbol{M}_{1,qd}^{ltp} + \boldsymbol{H}_i^{ltp}\boldsymbol{I}_{g,qd}^{ltp} + \boldsymbol{H}_i^{ltp}\boldsymbol{H}_v^{ltp}\boldsymbol{U}_{t,qd}^{ltp})\boldsymbol{G}_{tu}$$

Conversion to ABC stationary coordinate system gives:

$$\Delta \boldsymbol{m}_1^{ltp} = \boldsymbol{T}_{abc/dq}^{ltp(\theta)}\Delta \boldsymbol{m}_{1,dq}^{ltp} = \boldsymbol{T}_{gi}\Delta \boldsymbol{i}_g^{ltp} + \boldsymbol{T}_{gu}\Delta \boldsymbol{u}_g^{ltp}$$

$$\boldsymbol{T}_{gi} = \boldsymbol{T}_{abc/dq}^{ltp(\theta)}\boldsymbol{A}\boldsymbol{T}_{dq/abc}^{ltp(\theta)}, \boldsymbol{T}_{gu} = \boldsymbol{T}_{abc/dq}^{ltp(\theta)}\boldsymbol{B}\boldsymbol{T}_{dq/abc}^{ltp(\theta)} \tag{33}$$

Similarly, the double-frequency harmonic circulating current suppression control (CCSC) loop is derived as (34), where (35) is Toeplitz matrix of the current control block.

$$\left[\Delta \boldsymbol{u}_{c,dref}^{ltp}, \Delta \boldsymbol{u}_{c,qref}^{ltp}\right]^T = \boldsymbol{H}_c^{ltp}\left(-\Delta \boldsymbol{i}_{c,dq}^{ltp}\right) - \boldsymbol{T}_{pll\_c0}^{ltp}\cdot 2\Delta \boldsymbol{\theta}^{ltp}$$

$$\boldsymbol{T}_{pll\_c0}^{ltp} = -\boldsymbol{H}_c^{ltp}\boldsymbol{I}_{c,qd}^{ltp} + \boldsymbol{U}_{c,qd}^{ltp}, \boldsymbol{U}_{c,qd}^{ltp} = \left[\boldsymbol{U}_{cq0}^{ltp}, -\boldsymbol{U}_{cd0}^{ltp}\right]^T$$

$$\Delta \boldsymbol{m}_{2,dq}^{ltp} = \left[\Delta \boldsymbol{u}_{c,dref}^{ltp}, \Delta \boldsymbol{u}_{c,qref}^{ltp}\right]^T / U_{dc}, \boldsymbol{M}_{2,qd}^{ltp} = \boldsymbol{U}_{c,qd}^{ltp}/U_{dc} \tag{34}$$

$$\Delta \boldsymbol{i}_{c,dq}^{ltp} = \left[\Delta \boldsymbol{i}_{c,d}^{ltp}, \Delta \boldsymbol{i}_{c,q}^{ltp}\right]^T, \boldsymbol{I}_{c,qd}^{ltp} = \left[\boldsymbol{I}_{c,q}^{ltp}, -\boldsymbol{I}_{c,d}^{ltp}\right]^T$$

$$\boldsymbol{H}_c^{ltp} = \begin{bmatrix} \boldsymbol{G}_c^{ltp} & \boldsymbol{K}_c^{ltp} \\ -\boldsymbol{K}_c^{ltp} & \boldsymbol{G}_c^{ltp} \end{bmatrix}, \boldsymbol{G}_c^{ltp} = \text{diag}\left(k_{pic} + \frac{k_{iic}}{s+jk\omega_0}\right) \tag{35}$$

Conversion to ABC stationary coordinate system gives:

$$\Delta \boldsymbol{m}_2^{ltp} = \boldsymbol{T}_{abc/dq}^{ltp(2\theta)}\Delta \boldsymbol{m}_{2,dq}^{ltp} = \boldsymbol{T}_{cic}\Delta \boldsymbol{i}_c^{ltp} + \boldsymbol{T}_{cig}\Delta \boldsymbol{i}_g^{ltp} + \boldsymbol{T}_{cu}\Delta \boldsymbol{u}_c^{ltp}$$

$$\boldsymbol{T}_{cic} = -\boldsymbol{T}_{abc/dq}^{ltp(2\theta)}\boldsymbol{H}_c^{ltp}\boldsymbol{T}_{dq/abc}^{ltp(2\theta)}$$

$$\boldsymbol{T}_{cig} = 2\boldsymbol{T}_{abc/dq}^{ltp(2\theta)}(\boldsymbol{M}_{2,qd}^{ltp} + \boldsymbol{H}_c^{ltp}\boldsymbol{I}_{c,qd}^{ltp})\boldsymbol{G}_{ti}\boldsymbol{T}_{dq/abc}^{ltp(\theta)} \tag{36}$$

$$\boldsymbol{T}_{cu} = 2\boldsymbol{T}_{abc/dq}^{ltp(2\theta)}(\boldsymbol{M}_{2,qd}^{ltp} + \boldsymbol{H}_c^{ltp}\boldsymbol{I}_{c,qd}^{ltp})\boldsymbol{G}_{tu}\boldsymbol{T}_{dq/abc}^{ltp(\theta)}$$

The modulation index of the upper and lower arms are related to the differential-mode and common-mode circuit of MMC as (37) and (38), which is consistent with the general model (2).

$$\Delta \boldsymbol{m}^{\text{ltp}} = \begin{bmatrix} \Delta \boldsymbol{m}_{\text{u}}^{\text{ltp}} \\ \Delta \boldsymbol{m}_{\text{L}}^{\text{ltp}} \end{bmatrix} = \begin{bmatrix} -\Delta \boldsymbol{m}_1^{\text{ltp}} - \Delta \boldsymbol{m}_2^{\text{ltp}} \\ \Delta \boldsymbol{m}_1^{\text{ltp}} - \Delta \boldsymbol{m}_2^{\text{ltp}} \end{bmatrix} / 2 = \boldsymbol{F}\Delta \boldsymbol{i}^{\text{ltp}} + \boldsymbol{B}\Delta \boldsymbol{u}_{\text{g}}^{\text{ltp}} \quad (37)$$

$$\boldsymbol{F} = \begin{bmatrix} -\boldsymbol{T}_{\text{cic}} & -(\boldsymbol{T}_{\text{cig}} + \boldsymbol{T}_{\text{gi}}) \\ -\boldsymbol{T}_{\text{cic}} & -(\boldsymbol{T}_{\text{cig}} - \boldsymbol{T}_{\text{gi}}) \end{bmatrix} / 2, \boldsymbol{B} = \begin{bmatrix} -(\boldsymbol{T}_{\text{gu}} + \boldsymbol{T}_{\text{cu}}) \\ \boldsymbol{T}_{\text{gu}} - \boldsymbol{T}_{\text{cu}} \end{bmatrix} / 2 \quad (38)$$

### E. Complete Admittance Model

From the models of MMC main circuit and modulation index, the total admittance model is obtained, which is consistent with the general admittance model (3).

$$\Delta \boldsymbol{i}^{\text{ltp}} = -\left[ \boldsymbol{Z}_{\text{Larm}} + \boldsymbol{M}_{\text{v}}^{\text{ltp}} \boldsymbol{Z}_{\text{Ceq}} \boldsymbol{M}_{\text{i}}^{\text{ltp}} + (\boldsymbol{U}_{\text{C}}^{\text{ltp}} + \boldsymbol{M}_{\text{v}}^{\text{ltp}} \boldsymbol{Z}_{\text{Ceq}} \boldsymbol{I}_{\text{s}}^{\text{ltp}}) \boldsymbol{F} \right]^{-1}$$
$$\left[ \boldsymbol{E}_{\text{t}} + (\boldsymbol{U}_{\text{C}}^{\text{ltp}} + \boldsymbol{M}_{\text{v}}^{\text{ltp}} \boldsymbol{Z}_{\text{Ceq}} \boldsymbol{I}_{\text{s}}^{\text{ltp}}) \boldsymbol{B} \right] \Delta \boldsymbol{u}_{\text{g}}^{\text{ltp}} \equiv \boldsymbol{Y}_{\text{MMC}} \Delta \boldsymbol{u}_{\text{g}}^{\text{ltp}} \quad (39)$$

Since the current vector ($i$) contains both common-mode and differential-mode signals, while the concerned admittances correspond to the differential signals ($i_{\text{g}}$), the final admittance only needs to extract the corresponding rows and columns.

## IV. ADMITTANCE MODELING OF GRID-FOLLOWING MMC WITH CONSTANT DC VOLTAGE CONTROL

The outer DC voltage control loop and inner control loop are sketched in Fig. 4(a) and Fig. 4(b), respectively. Phase locked loop (PLL) is used to obtain the grid voltage, frequency and phase angle for synchronization.

### A. Modeling of MMC Main Circuit

In this case, the DC current is constant while DC voltage is to be regulated. LTP model of MMC bridge arm capacitor voltage is the same as (4). Considering the DC voltage dynamics, LTP model of MMC arm current is given by (40).

$$\boldsymbol{Z}_{\text{Larm}} \Delta \boldsymbol{i}^{\text{ltp}} = -(\boldsymbol{E}_{\text{t}} \Delta \boldsymbol{u}_{\text{g}}^{\text{ltp}} + \boldsymbol{U}_{\text{C}}^{\text{ltp}} \Delta \boldsymbol{m}^{\text{ltp}} + \boldsymbol{M}_{\text{v}}^{\text{ltp}} \Delta \boldsymbol{u}_{\text{C}}^{\text{ltp}} + \boldsymbol{E}_{\text{v}} \Delta \boldsymbol{u}_{\text{dc}}^{\text{ltp}})$$
$$\boldsymbol{E}_{\text{v}} = \left[ \boldsymbol{I}_{3(2h+1)}, \boldsymbol{0} \right]^T \quad (40)$$

To establish a unified impedance model, it is necessary to express DC voltage in terms of AC-side variables. From power balance between the AC and DC sides of MMC, we have (41), where $P_{\text{dc}}$ is the output power from DC side, $P_{\text{Cu}}$, $P_{\text{Cl}}$, $P_{\text{Lu}}$, $P_{\text{Ll}}$ denote the transient power of capacitors and inductors in the upper and lower arms. $P_{\text{out}}$ represents the output active power. The detailed relation in frequency domain is given by (42).

$$P_{\text{dc}} = P_{\text{Cu}} + P_{\text{Cl}} + P_{\text{Lu}} + P_{\text{Ll}} + P_{\text{out}} \quad (41)$$

$$U_{\text{dc}} \Delta \boldsymbol{i}_{\text{dc}}^{\text{ltp}} + I_{\text{dc}} \Delta \boldsymbol{u}_{\text{dc}}^{\text{ltp}} = \boldsymbol{U}_{\text{g}}^{\text{ltp}} \Delta \boldsymbol{i}_{\text{g}}^{\text{ltp}} + \boldsymbol{I}_{\text{g}}^{\text{ltp}} \Delta \boldsymbol{u}_{\text{g}}^{\text{ltp}} +$$
$$(\boldsymbol{Y}_{\text{C}} \boldsymbol{U}_{\text{Cu}} + \boldsymbol{U}_{\text{Cu}} \boldsymbol{Y}_{\text{C}})/2 \cdot \Delta \boldsymbol{u}_{\text{Cu}}^{\text{ltp}} + (\boldsymbol{Y}_{\text{C}} \boldsymbol{U}_{\text{Cl}} + \boldsymbol{U}_{\text{Cl}} \boldsymbol{Y}_{\text{C}})/2 \cdot \Delta \boldsymbol{u}_{\text{Cl}}^{\text{ltp}} \quad (42)$$
$$+ (\boldsymbol{Z}_{\text{L}} \boldsymbol{I}_{\text{Lu}} + \boldsymbol{I}_{\text{Lu}} \boldsymbol{Z}_{\text{L}})/2 \cdot \Delta \boldsymbol{i}_{\text{Lu}}^{\text{ltp}} + (\boldsymbol{Z}_{\text{L}} \boldsymbol{I}_{\text{Ll}} + \boldsymbol{I}_{\text{Ll}} \boldsymbol{Z}_{\text{L}})/2 \cdot \Delta \boldsymbol{i}_{\text{Ll}}^{\text{ltp}}$$

Further, considering the arm current relation (43) and $\Delta \boldsymbol{i}_{\text{dc}}^{\text{hss}} = 0$, the power relation is further expressed as (44).

$$\Delta \boldsymbol{i}_{\text{Lu}}^{\text{ltp}} = \Delta \boldsymbol{i}_{\text{c}}^{\text{ltp}} + \Delta \boldsymbol{i}_{\text{g}}^{\text{ltp}}/2, \Delta \boldsymbol{i}_{\text{Ll}}^{\text{ltp}} = \Delta \boldsymbol{i}_{\text{c}}^{\text{ltp}} - \Delta \boldsymbol{i}_{\text{g}}^{\text{ltp}}/2 \quad (43)$$

$$I_{\text{dc}} \Delta \boldsymbol{u}_{\text{dc}}^{\text{ltp}} = (\boldsymbol{Z}_{\text{L}} \boldsymbol{I}_{\text{Lu}} + \boldsymbol{I}_{\text{Lu}} \boldsymbol{Z}_{\text{L}} - \boldsymbol{Z}_{\text{L}} \boldsymbol{I}_{\text{Ll}} - \boldsymbol{I}_{\text{Ll}} \boldsymbol{Z}_{\text{L}} + 4\boldsymbol{U}_{\text{g}})/4 \cdot \Delta \boldsymbol{i}_{\text{g}}^{\text{ltp}}$$
$$+ (\boldsymbol{Z}_{\text{L}} \boldsymbol{I}_{\text{Lu}} + \boldsymbol{I}_{\text{Lu}} \boldsymbol{Z}_{\text{L}} + \boldsymbol{Z}_{\text{L}} \boldsymbol{I}_{\text{Ll}} + \boldsymbol{I}_{\text{Ll}} \boldsymbol{Z}_{\text{L}})/2 \cdot \Delta \boldsymbol{i}_{\text{c}}^{\text{ltp}} + \boldsymbol{I}_{\text{g}} \Delta \boldsymbol{u}_{\text{g}}^{\text{ltp}} \quad (44)$$
$$+ (\boldsymbol{Y}_{\text{C}} \boldsymbol{U}_{\text{Cu}} + \boldsymbol{U}_{\text{Cu}} \boldsymbol{Y}_{\text{C}})/2 \cdot \Delta \boldsymbol{u}_{\text{Cu}}^{\text{ltp}} + (\boldsymbol{Y}_{\text{C}} \boldsymbol{U}_{\text{Cl}} + \boldsymbol{U}_{\text{Cl}} \boldsymbol{Y}_{\text{C}})/2 \cdot \Delta \boldsymbol{u}_{\text{Cl}}^{\text{ltp}}$$

The above equation can be formulated by (45), and the model of DC voltage dynamics can then be rewritten as (46).

$$\Delta \boldsymbol{u}_{\text{dc}}^{\text{ltp}} = \boldsymbol{f}_1 \Delta \boldsymbol{i}_{\text{c}}^{\text{ltp}} + \boldsymbol{f}_2 \Delta \boldsymbol{u}_{\text{Cu}}^{\text{ltp}} + \boldsymbol{f}_3 \Delta \boldsymbol{u}_{\text{Cl}}^{\text{ltp}} + \boldsymbol{f}_4 \Delta \boldsymbol{i}_{\text{g}}^{\text{ltp}} + \boldsymbol{f}_5 \Delta \boldsymbol{u}_{\text{g}}^{\text{ltp}}$$
$$\boldsymbol{f}_1 = (\boldsymbol{Z}_{\text{L}} \boldsymbol{I}_{\text{Lu}} + \boldsymbol{I}_{\text{Lu}} \boldsymbol{Z}_{\text{L}} + \boldsymbol{Z}_{\text{L}} \boldsymbol{I}_{\text{Ll}} + \boldsymbol{I}_{\text{Ll}} \boldsymbol{Z}_{\text{L}})/(2I_{\text{dc}})$$
$$\boldsymbol{f}_2 = (\boldsymbol{Y}_{\text{C}} \boldsymbol{U}_{\text{Cu}} + \boldsymbol{U}_{\text{Cu}} \boldsymbol{Y}_{\text{C}})/(2I_{\text{dc}}), \boldsymbol{f}_3 = (\boldsymbol{Y}_{\text{C}} \boldsymbol{U}_{\text{Cl}} + \boldsymbol{U}_{\text{Cl}} \boldsymbol{Y}_{\text{C}})/(2I_{\text{dc}}) \quad (45)$$
$$\boldsymbol{f}_4 = (\boldsymbol{Z}_{\text{L}} \boldsymbol{I}_{\text{Lu}} + \boldsymbol{I}_{\text{Lu}} \boldsymbol{Z}_{\text{L}} + \boldsymbol{Z}_{\text{L}} \boldsymbol{I}_{\text{Ll}} + \boldsymbol{I}_{\text{Ll}} \boldsymbol{Z}_{\text{L}} - 4\boldsymbol{U}_{\text{g}})/(4I_{\text{dc}})$$
$$\boldsymbol{f}_5 = \boldsymbol{I}_{\text{g}}/I_{\text{dc}}$$

$$\Delta \boldsymbol{u}_{\text{dc}}^{\text{ltp}} = \boldsymbol{K}_{\text{vi}} \Delta \boldsymbol{i}^{\text{ltp}} + \boldsymbol{K}_{\text{uc}} \Delta \boldsymbol{u}_{\text{C}}^{\text{ltp}} + \boldsymbol{K}_{\text{ug}} \Delta \boldsymbol{u}_{\text{g}}^{\text{ltp}}$$
$$\boldsymbol{K}_{\text{vi}} = [\boldsymbol{f}_1, \boldsymbol{f}_4], \boldsymbol{K}_{\text{uc}} = [\boldsymbol{f}_2, \boldsymbol{f}_3], \boldsymbol{K}_{\text{ug}} = \boldsymbol{f}_5 \quad (46)$$

Substituting the DC voltage model into MMC main circuit model (40) yields (47), which is consistent with the general model (1). The developed model is different from MMC circuit model (13) where DC voltage is constant.

$$(\boldsymbol{Z}_{\text{Larm}} + \boldsymbol{E}_{\text{v}} \boldsymbol{K}_{\text{vi}} + (\boldsymbol{M}_{\text{v}}^{\text{ltp}} + \boldsymbol{E}_{\text{v}} \boldsymbol{K}_{\text{uc}}) \boldsymbol{Z}_{\text{Ceq}} \boldsymbol{M}_{\text{i}}^{\text{ltp}}) \Delta \boldsymbol{i}^{\text{ltp}} =$$
$$-((\boldsymbol{E}_{\text{t}} + \boldsymbol{E}_{\text{v}} \boldsymbol{K}_{\text{ug}}) \Delta \boldsymbol{u}_{\text{g}}^{\text{ltp}} + \left[ \boldsymbol{U}_{\text{C}}^{\text{ltp}} + (\boldsymbol{M}_{\text{v}}^{\text{ltp}} + \boldsymbol{E}_{\text{v}} \boldsymbol{K}_{\text{uc}}) \boldsymbol{Z}_{\text{Ceq}} \boldsymbol{I}_{\text{s}}^{\text{ltp}} \right] \Delta \boldsymbol{m}^{\text{ltp}}) \quad (47)$$

### B. Coordinate Transformation and Synchronization

The modeling of coordinate transformation is the same as (26) in the grid-forming case. Based on three-phase Park transform at fundamental frequency, LTP model of PLL is derived as (48), where (49) is obtained by frequency shifting the PI control $H_{\text{pll}}(s) = (k_{\text{p,PLL}} + k_{\text{i,PLL}}/s)/s$ within PLL.

$$\Delta \boldsymbol{\theta}_{\text{pll}}^{\text{ltp}} = \boldsymbol{T}_{\text{pll}}^{\text{ltp}} \boldsymbol{T}_{\text{sin}}^{\text{ltp}} \cdot \Delta \boldsymbol{u}_{\text{g}}^{\text{ltp}}, \boldsymbol{T}_{\text{pll}}^{\text{ltp}} = \frac{\boldsymbol{H}_{\text{pll}}^{\text{ltp}}}{\boldsymbol{I} + \boldsymbol{U}_{\text{gd}}^{\text{ltp}} \cdot \boldsymbol{H}_{\text{pll}}^{\text{ltp}}} \quad (48)$$

$$\boldsymbol{H}_{\text{pll}}^{\text{ltp}} = \text{diag}\left( H_{\text{pll}}(s - jh\omega_1), \cdots, H_{\text{pll}}(s), \cdots, H_{\text{pll}}(s + jh\omega_1) \right) \quad (49)$$

### C. Control Part

LTP model of DC voltage control loop is given by (50), where $\boldsymbol{H}_{\text{d}}^{\text{ltp}}$ is obtained by frequency shifting and matrix expansion of the control block $H_{\text{d}} = k_{\text{pd}} + k_{\text{id}}/s$.

$$\begin{bmatrix} \Delta \boldsymbol{i}_{\text{g,dref}}^{\text{ltp}} \\ \Delta \boldsymbol{i}_{\text{g,qref}}^{\text{ltp}} \end{bmatrix} = \begin{bmatrix} \boldsymbol{H}_{\text{d}}^{\text{ltp}} \\ \boldsymbol{0} \end{bmatrix} \Delta \boldsymbol{u}_{\text{dc}}^{\text{ltp}} \quad (50)$$

Combining PLL model (48), LTP model of the fundamental current control loop is then given by (51), where $H_i^{ltp}$ is obtained by frequency shifting of the current control block $H_i = k_{pi} + k_{ii}/s$, and the cross-coupling term is ignored for simplicity but without loss of generality.

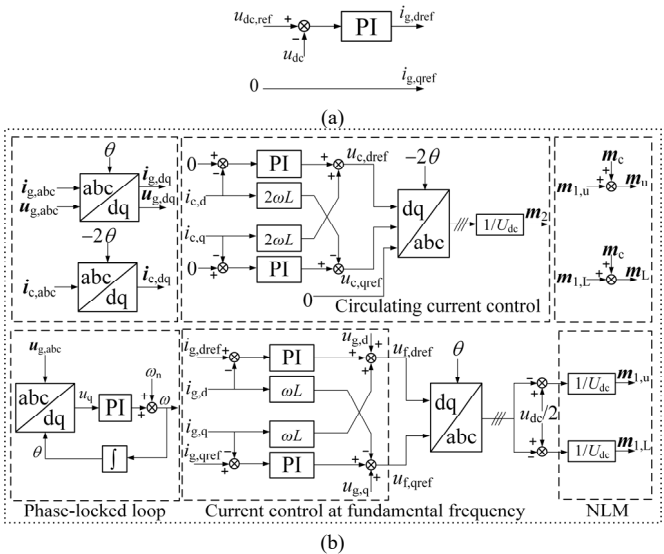

Fig. 4. Diagram of MMC DC voltage control. (a) Outer loop. (b) Inner loop.

$$\begin{bmatrix} \Delta u_{f,dref}^{ltp} \\ \Delta u_{f,qref}^{ltp} \end{bmatrix} = -\begin{bmatrix} H_i^{ltp} H_d \\ 0 \end{bmatrix} \Delta u_{dc}^{ltp} - G_i^{ltp} T_{abc/dq}^{ltp(\theta)} \Delta i_g^{ltp} - T_{pll,g0}^{ltp} T_{pll}^{ltp} T_{sin}^{ltp} \cdot \Delta u_g^{ltp} \quad (51)$$

Applying fundamental-frequency Park transformation (26), the fundamental modulation index is obtained as (52).

$$\Delta u_{f,ref}^{ltp} = T_{dq/abc}^{ltp(\theta)} \left[ \Delta u_{f,dref}^{ltp}, \Delta u_{f,qref}^{ltp} \right]^T \quad (52)$$
$$\Delta m_1^{ltp} = \Delta u_f^{ltp}/U_{dc} \equiv T_{gv} \Delta u_{dc}^{ltp} + T_{gi} \Delta i_g^{ltp} + T_{gu} \Delta u_g^{ltp}$$

Similarly, considering the coordinate transformation model (26) in double-frequency coordinate system, the corresponding LTP model of CCSC is similar to the grid-forming case, with phase angle changed to angle from PLL.

$$\left[ \Delta u_{c,dref}^{ltp}, \Delta u_{c,qref}^{ltp} \right]^T = H_c^{ltp} \left( -\Delta i_{c,dq}^{ltp} \right) - T_{pll\_c0}^{ltp} \cdot 2\Delta \theta_{pll}^{ltp} \quad (53)$$

$$\Delta u_{c,ref}^{ltp} = -T_{dq/abc}^{ltp(2\theta)} G_c^{ltp} T_{abc/dq}^{ltp(2\theta)} \Delta i_c^{ltp} - 2 T_{dq/abc}^{ltp(2\theta)} T_{pll\_c0}^{ltp} T_{pll}^{ltp} T_{sin}^{ltp} \Delta u_g^{ltp} \quad (54)$$

$$\Delta m_2^{ltp} = \Delta u_c^{ltp*}/U_{dc} \equiv T_{cic} \Delta i_c^{ltp} + T_{cu} \Delta u_g^{ltp} \quad (55)$$

### D. Modulation Index Generation

The modulation indices of the upper and lower bridge arms of MMC are associated with fundamental differential-mode circuit and common-mode circuit by:

$$\Delta m^{ltp} = \begin{bmatrix} \Delta m_u^{ltp} \\ \Delta m_L^{ltp} \end{bmatrix} = \begin{bmatrix} -\Delta m_1^{ltp} - \Delta m_2^{ltp} \\ \Delta m_1^{ltp} - \Delta m_2^{ltp} \end{bmatrix}/2 = \frac{1}{2}\begin{bmatrix} -T_{gv} \\ T_{gv} \end{bmatrix} \Delta u_{dc}^{ltp}$$
$$+ \frac{1}{2}\begin{bmatrix} -T_{cic} & -T_{gi} \\ -T_{cic} & T_{gi} \end{bmatrix} \Delta i^{ltp} + \frac{1}{2}\begin{bmatrix} -(T_{cu}+T_{gu}) \\ -(T_{cu}-T_{gu}) \end{bmatrix} \Delta u_g^{ltp} \quad (56)$$

Considering DC voltage model (46), the total modulation index of MMC can be rewritten as (57)-(59).

$$\Delta m^{ltp} = J_i \Delta i^{ltp} + J_u \Delta u_g^{ltp} + J_v \Delta u_C^{ltp} \quad (57)$$

$$J_i = \frac{1}{2}\begin{bmatrix} -T_{cic} & -T_{gi} \\ -T_{cic} & T_{gi} \end{bmatrix} + \frac{1}{2}\begin{bmatrix} -T_{gv} \\ T_{gv} \end{bmatrix} K_{vi} \quad (58)$$

$$J_u = \frac{1}{2}\begin{bmatrix} -(T_{cu}+T_{gu}) \\ -(T_{cu}-T_{gu}) \end{bmatrix} + \frac{1}{2}\begin{bmatrix} -T_{gv} \\ T_{gv} \end{bmatrix} K_{ug}, J_v = \frac{1}{2}\begin{bmatrix} -T_{gv} \\ T_{gv} \end{bmatrix} K_{uc} \quad (59)$$

Further substituting the capacitor model (4) into (57) gives:

$$\Delta m^{ltp} = F \Delta i^{ltp} + B \Delta u_g^{ltp} \quad (60)$$

$$F = \left[ E_t - J_v Z_{Ceq} I_s^{ltp} \right]^{-1} (J_i + J_v M_i^{ltp})$$
$$B = \left[ E_t - J_v Z_{Ceq} I_s^{ltp} \right]^{-1} J_u \quad (61)$$

### E. Complete Admittance Model

Combining MMC circuit model (47) and modulation index model (60), the complete admittance model of MMC and control system is obtained as (62), which is different from (39) due to the regulated DC voltage.

$$-\begin{pmatrix} Z_{Larm} + E_v K_{vi} + (M_v^{ltp} + E_v K_{uc}) Z_{Ceq} M_i^{ltp} + \\ \left[ U_C^{ltp} + (M_v^{ltp} + E_v K_{uc}) Z_{Ceq} I_s^{ltp} \right] \cdot F \end{pmatrix} \Delta i^{ltp}$$
$$= \begin{pmatrix} (E_t + E_v K_{ug}) + \\ \left[ U_C^{ltp} + (M_v^{ltp} + E_v K_{uc}) Z_{Ceq} I_s^{ltp} \right] \cdot B \end{pmatrix} \Delta u_g^{ltp} \quad (62)$$

## V. EXTENSION TO OTHER CONTROL AND CONVERTER TYPES

The general modeling procedure in (1)-(3) can be extended to other control types and converter topologies. This section gives examples of MMC with grid-following PQ control and 2L-VSC with grid-forming control, and makes a general summary afterwards.

### A. Admittance of MMC with Closed-loop PQ Control

#### 1) The Power Control Loop

The outer power control loop is shown in Fig. 5, which can also cover constant current control with the outer control blocks being zero. The small-signal model of MMC current reference value is given by (63), where the transfer function of the power control module is given by (64).

$$\Delta i_{gref,dq}^{ltp} = \begin{bmatrix} \Delta i_{g,dref}^{ltp} \\ \Delta i_{g,qref}^{ltp} \end{bmatrix} = H_p^{ltp} U_{PQ0} \Delta i_{g,dq}^{ltp} + H_p^{ltp} I_{PQ0} \Delta u_{g,dq}^{ltp}$$
$$U_{PQ0} = \frac{3}{2}\begin{bmatrix} U_{g,d0}^{ltp} & U_{g,q0}^{ltp} \\ U_{g,q0}^{ltp} & -U_{g,d0}^{ltp} \end{bmatrix}, I_{PQ0} = \frac{3}{2}\begin{bmatrix} I_{g,d0}^{ltp} & I_{g,q0}^{ltp} \\ -I_{g,q0}^{ltp} & I_{g,d0}^{ltp} \end{bmatrix} \quad (63)$$

$$H_p^{ltp} = \text{blkdiag}\left( G_p^{ltp}, -G_p^{ltp} \right), G_p^{ltp} = \text{diag}\left( k_{pp} + \frac{k_{ip}}{s + jk\omega_0} \right) \quad (64)$$

#### 2) Fundamental Current Control Loop

Similar to grid-forming control and considering coordinate transformation, the model of fundamental current control loop is derived as:

$$\begin{bmatrix} \Delta u_{f,dref}^{ltp} \\ \Delta u_{f,qref}^{ltp} \end{bmatrix} = \left( H_p^{ltp} H_i^{ltp} U_{PQ0} - \begin{bmatrix} H_i^{ltp} & -K_i^{ltp} \\ K_i^{ltp} & H_i^{ltp} \end{bmatrix} \right) T_{abc/dq}^{ltp(\theta)} \Delta i_g^{ltp}$$
$$+ \left\{ \begin{bmatrix} H_i^{ltp} & \\ & H_i^{ltp} \end{bmatrix} I_{PQ0} T_{abc/dq}^{ltp(\theta)} - T_{pll,g0}^{ltp} \right\} \cdot \Delta u_g^{ltp} \quad (65)$$

$$T_{pll,g0}^{ltp} = \left( \begin{bmatrix} H_i^{ltp} & -K_i^{ltp} \\ K_i^{ltp} & H_i^{ltp} \end{bmatrix}\begin{bmatrix} I_{gq0}^{ltp} \\ -I_{gd0}^{ltp} \end{bmatrix} + \begin{bmatrix} U_{fq0}^{ltp} \\ -U_{fd0}^{ltp} \end{bmatrix} \right) T_{pll}^{ltp} T_{sin}^{ltp}$$

$$\Delta u_f^{ltp} = T_{dq/abc}^{ltp(\theta)} \left[ \Delta u_{f,dref}^{ltp}, \Delta u_{f,qref}^{ltp} \right]^T \quad (66)$$
$$\Delta m_1^{ltp} = \Delta u_f^{ltp}/U_{dc} \equiv T_{gi} \cdot \Delta i_g^{ltp} + T_{gu} \cdot \Delta u_g^{ltp}$$

#### 3) Modulation Index and Complete Admittance Model

Combining CCSC model (55), model of the total modulation index is similar to (37), with relevant variables given by (67). The subsequent steps and the total impedance expression are the same as the grid-forming case.

$$G_i = \begin{bmatrix} -T_{cic} & -T_{gi} \\ -T_{cic} & T_{gi} \end{bmatrix}/2, G_u = \begin{bmatrix} -(T_{gu}+T_{cu}) \\ T_{gu}-T_{cu} \end{bmatrix}/2 \quad (67)$$

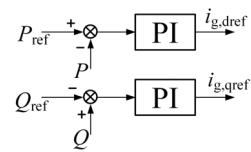

Fig. 5. Diagram of outer power control loop.

## B. Admittance of 2L-VSC with Grid-forming Control

Since DC voltage is constant and there is no arm capacitor, LTP model of the 2-level VSC main circuit can be directly modeled as the filter inductor in (68). A L-filter is considered here, but can be extended to LC filter as well.

$$\boldsymbol{Z}_L \Delta \boldsymbol{i}^{ltp} = -(\boldsymbol{E}_t \Delta \boldsymbol{u}_g^{ltp} + \boldsymbol{U}_{dc}^{ltp} \Delta \boldsymbol{m}^{ltp}) \quad (68)$$

The control part generates the modulation index which is modeled similar to the fundamental current control of MMC:

$$\Delta \boldsymbol{m}^{ltp} = \boldsymbol{G}_i \Delta \boldsymbol{i}^{ltp} + \boldsymbol{G}_u \Delta \boldsymbol{u}_g^{ltp} \quad (69)$$

The final impedance model is then derived as:

$$\Delta \boldsymbol{i}^{ltp} = -\left(\boldsymbol{Z}_L + \boldsymbol{U}_{dc}^{ltp} \boldsymbol{G}_i\right)^{-1} \left(\boldsymbol{E}_t + \boldsymbol{U}_{dc}^{ltp} \boldsymbol{G}_u\right) \boldsymbol{u}_g^{ltp} \equiv \boldsymbol{Y}_{2L} \boldsymbol{u}_g^{ltp} \quad (70)$$

It can be seen the admittance of 2L-VSC can be obtained by removing the arm capacitance term $\boldsymbol{Z}_{Ceq}$ from MMC admittance (39), indicating unification of the impedance model of MMC and 2-level VSC. This is applicable regardless of the specific control types adopted.

## C. Summary of Converter Admittance Models

The impedance models established above can be categorized into the general forms in (1)-(3). The formulations of modulation index are uniform, but the detailed expressions $\boldsymbol{B}$ and $\boldsymbol{F}$ depend on the specific control strategies. These are also the core factors that distinguish the impedance for different control types. A summary of the admittance models for MMC and 2L-VSC with different control types are given in Table I.

From the explicit expressions of MMC admittance, some insightful inferences can be made. For example, $\boldsymbol{Z}_{Ceq}$ tends to zero in high frequency, which will make MMC admittance model similar to 2L-VSC counterpart, while the main difference occurs in the low-to-middle frequency range where $\boldsymbol{Z}_{Ceq}$ cannot be neglected. For the same reason, as the submodule capacitance increases, $\boldsymbol{Z}_{Ceq}$ will decrease, and MMC impedance model will become increasingly closer to the admittance model of 2L-VSC. In cases where large capacitors or batteries are embedded in the submodules, the impedance of MMC can be rapidly estimated from the 2L-VSC counterpart, especially when the sensor and modulation delays are not concerned.

## VI. IMPEDANCE MEASUREMENT AND VERIFICATION

Grid-connected MMC with parameters listed in Table II is simulated. Frequency-scanning method is adopted to measure the output admittance by injecting positive and negative sequence voltage harmonics at the terminal of MMC, and then conducting frequency spectrum analysis of the response current [17, 28]. The highest harmonic order in the admittance modeling is $h$=3, which is sufficient to ensure acceptable accuracy. Admittance elements in the "1" and "2" sequences are extracted from the obtained matrix [12].

Table I
Summary of admittance models for different converter and control types

| Converter /control types | $\boldsymbol{Z}$ | $\boldsymbol{C}$ | $\boldsymbol{D}$ |
|---|---|---|---|
| MMC GFM /GFL PQ | $\boldsymbol{Z}_{Larm} + \boldsymbol{M}_v^{ltp} \boldsymbol{Z}_{Ceq} \boldsymbol{M}_i^{ltp}$ | $-\boldsymbol{U}_C^{ltp} - \boldsymbol{M}_v^{ltp} \boldsymbol{Z}_{Ceq} \boldsymbol{I}_s^{ltp}$ | $-\boldsymbol{E}_t$ |
| MMC GFL DC voltage | $\boldsymbol{Z}_{Larm} + \boldsymbol{E}_v \boldsymbol{K}_{vi} + (\boldsymbol{M}_v^{ltp} + \boldsymbol{E}_v \boldsymbol{K}_{uc}) \boldsymbol{Z}_{Ceq} \boldsymbol{M}_i^{ltp}$ | $(\boldsymbol{M}_v^{ltp} + \boldsymbol{E}_v \boldsymbol{K}_{uc}) \boldsymbol{Z}_{Ceq} \boldsymbol{I}_s^{ltp} + \boldsymbol{U}_C^{ltp}$ | $-\boldsymbol{E}_t - \boldsymbol{E}_v \boldsymbol{K}_{ug}$ |
| 2L-VSC GFL PQ/GFM | $\boldsymbol{Z}_L$ | $-\boldsymbol{U}_{dc}^{ltp}$ | $-\boldsymbol{E}_t$ |

TABLE II
Parameters of the grid-connected MMC circuit

| Parameter | Value | Parameter | Value |
|---|---|---|---|
| $S_{rating}$ | 220 kVA | $U_{dc}$ | 135 kV |
| $\omega_0$ | 100 π rad/s | $U_{g(LL-RMS)}$ | 66 kV |
| $N$ | 50 | $L_{arm}$ | 16.37 mH |
| $C_m$ | 10.48 mF | $R_{arm}$ | 0.03 Ω |

### A. Admittance of MMC Circuit under Open-loop Control

The admittance of MMC main circuit in Table II, with constant DC voltage and no grid impedance, is tested. The modulation index exerted upon the fundamental current circuit is set as $m_d+jm_q$=0.8+j0.04, while the modulation index of circulating current circuit is zero, which are close to the close-loop grid-following control case. The accurately derived model (13) closely matches the measured admittance, as shown in Fig.

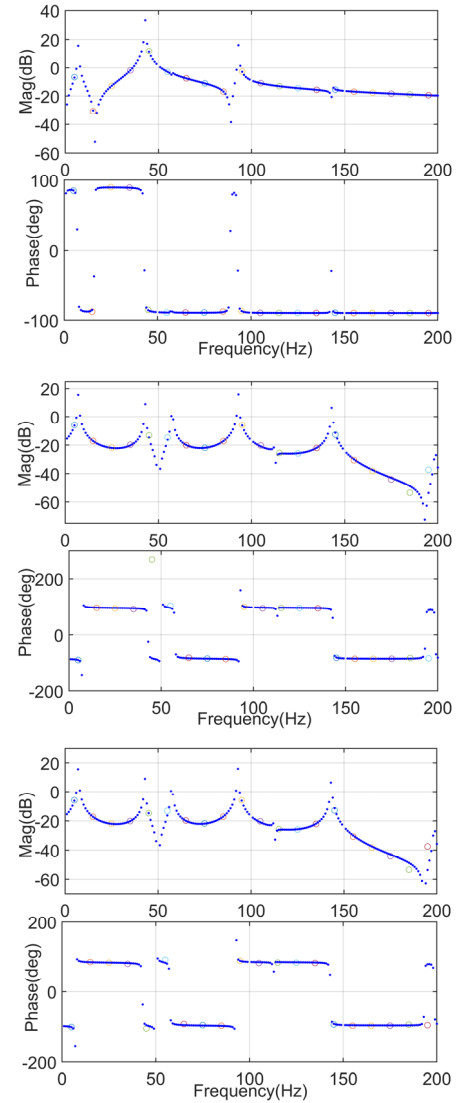

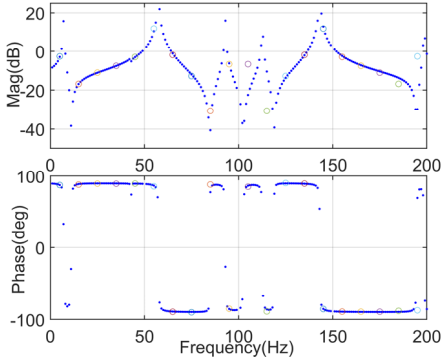

Fig. 6. Derived (blue dots) and measured (circles) admittance of MMC with open-loop control (Up to down: $Y_{11}$, $Y_{12}$, $Y_{21}$, $Y_{22}$).

6. Meanwhile, the curves of $Y_{11}$ and $Y_{22}$, $Y_{12}$ and $Y_{21}$ satisfy the general relation (71). Therefore, the four elements are not independent, with the first row alone sufficient to characterize the whole matrix [9]. Therefore, for brevity, only $Y_{11}$ and $Y_{12}$ are presented in the following results.

$$Y_{11}(s) = Y_{22}[(2j\omega_0 - s)]^* \\ Y_{12}(s) = Y_{21}[(2j\omega_0 - s)]^* \quad (71)$$

### B. Admittance of MMC with Grid-following Control

For MMC with closed-loop DC voltage and PQ control and parameters given in Table III, the measured admittances are close to the theoretically derived models (62) and (39), as illustrated in Fig. 7 and Fig. 8, respectively. The differences of these admittances are mainly in the lower-frequency range (<100 Hz), while close in the higher-frequency range (>100 Hz). This is reasonable, since the difference of MMC system are mainly the outer control loop in low-frequency range, while similar in high-frequency range where the physical parameters and the inner current control of MMC are dominant.

As seen, there are valley values at 50 Hz for grid-following control, which is consistent with the converter regulated as a current source with high parallel impedance (low admittance) at this frequency. As seen from Fig. 7, the sequence coupling is more severe for MMC with DC voltage control than the case with constant DC voltage source. This is reasonable since DC voltage control will introduce more asymmetry in addition to the existing unsymmetrical blocks such as PLL.

Fig. 8 also shows that frequencies at the valleys and peaks on the admittance of MMC with PQ control are close to MMC under open-loop control in Fig. 6, which can be further studied separately.

TABLE III
Parameters of MMC with different control strategies

| Parameter | Value | Parameter | Value |
|---|---|---|---|
| $k_{p,PLL}$ | 1800 p.u. | $k_{i,PLL}$ | 3200 p.u. |
| $k_{pd}$ | 4 p.u. | $k_{id}$ | 75 p.u. |
| $k_{pp}$ | 0.2 p.u. | $k_{ip}$ | 15 p.u. |
| $k_{pv}$ | 0.2 p.u. | $k_{iv}$ | 26 p.u. |
| $H$ | 1 s | $D$ | 100 p.u. |
| $D_v$ | 0.1 p.u.s | $T_v$ | 0.01 s |
| $k_{pi}$ | 2 p.u. | $k_{ii}$ | 100 p.u. |
| $k_{pic}$ | 1 p.u. | $k_{iic}$ | 5 p.u. |

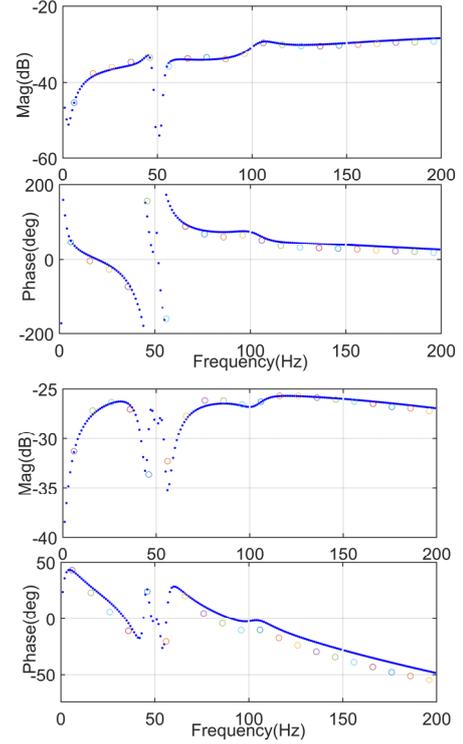

Fig. 7. Derived and measured admittance of MMC with DC voltage control (Up to down: $Y_{11}$, $Y_{12}$).

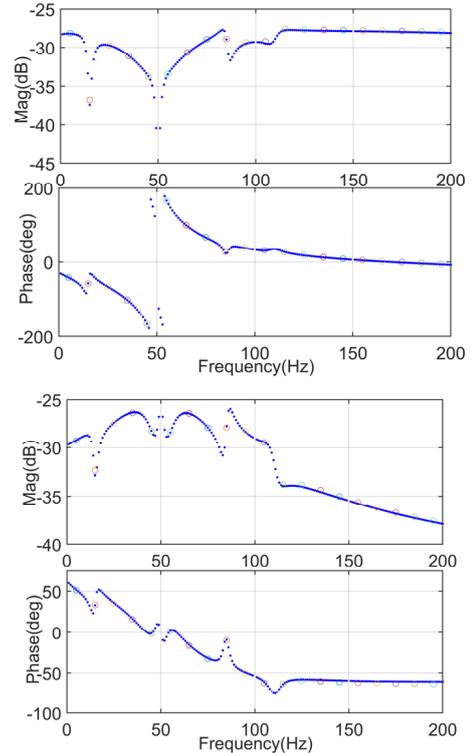

Fig. 8. Derived and measured admittance of MMC with closed-loop power control (Up to down: $Y_{11}$, $Y_{12}$).

### C. Admittance of MMC with Grid-forming Control and Comparison with 2L-VSC

To mitigate oscillation when MMC is connected with grid and thus reduce the time needed for frequency scanning, a grid resistor $R_g$=5 Ω is added, which is also included in the admittance model. The corresponding admittance under open-loop control ($m_d+jm_q$=0.84+j0.02) is developed and verified, as shown in Fig. 9. As seen from the phase plot, the admittance becomes more resistive, compared to the alternation between being inductive and capacitive in Fig. 6.

The admittance model (39) of MMC under grid-forming control is verified by comparing with the measured admittance. As observed in Fig. 10, there are peak values at 50 Hz, which is reasonable since the converter is controlled as a voltage source with low series impedance at this frequency. The admittance is then compared with the smooth admittance of 2L-VSC. As shown in Fig. 11, the frequencies at peaks and valleys of MMC can be found on the admittance curves under open-loop control in Fig. 9.

In Fig. 11, as the submodule capacitance increases to 10 times, the admittance of MMC under grid-forming control approaches that of 2L-VSC. The observation is also consistent with the analysis in Section VI. B.

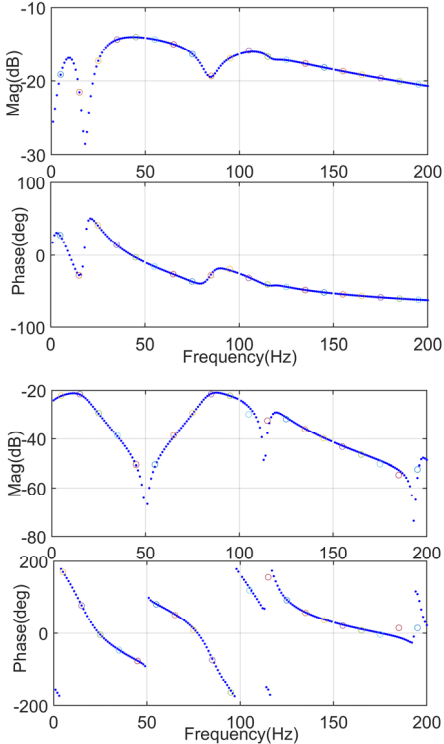

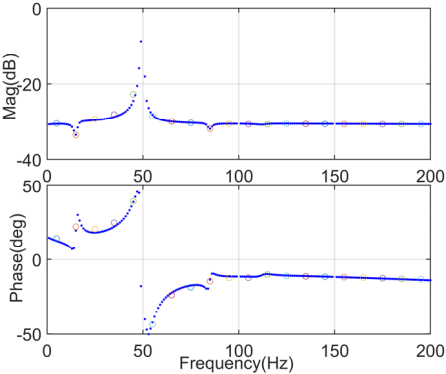

Fig. 9. Derived and measured admittance of MMC with open-loop control and added grid side rgesistor (Up to down: $Y_{11}$, $Y_{12}$).

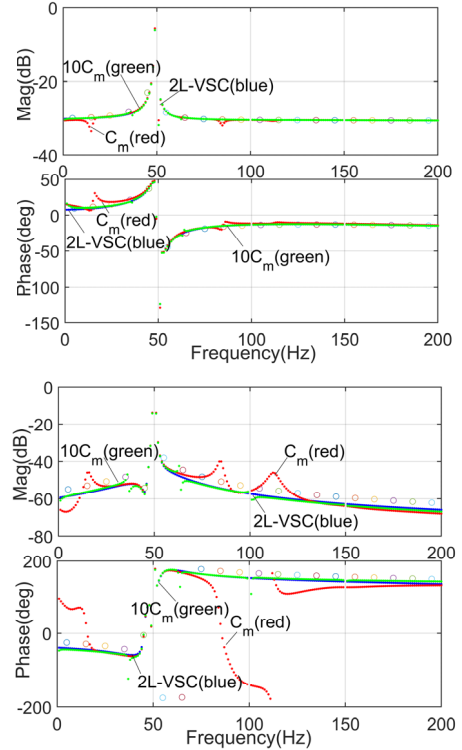

Fig. 10. Derived and measured admittance of MMC with grid-forming control (Up to down: $Y_{11}$, $Y_{12}$).

Fig. 11. Admittance of 2L-VSC and MMC with grid-forming control for different submodule capacitances (Up to down: $Y_{11}$, $Y_{12}$).

## VII. CONCLUSION

Linear time periodic theory is demonstrated as a powerful tool for modeling and analyzing complex power converters such as MMC in modern power system. Based on the multiple impedance models established and comparative analysis in this article, the following conclusions can be drawn.

(1) With the proposed general modeling procedure which interconnects the separately modeled circuit and control part, impedance models of MMC or other converters with different grid-forming and grid-following control are established.

(2) The developed impedance models have explicit form, based on which the influence of each part can be analyzed. It can be easily implied from the model that, with the increased submodule capacitance, MMC impedance will become closer to the two-level converter counterpart.

The general modeling procedure and the established MMC impedance model can pave foundations for further impedance stability research, including stability analysis and damping design, sensitivity analysis to analyze the impact of each part quantitively, and modeling of power converter system with the explicit impedance of each converter derived.